\documentclass[aps,pre,twocolumn,showpacs,floatfix,superscriptaddress,graphics,nofootinbib,longbibliography]{revtex4-1}

\usepackage{amsmath,amssymb,graphicx,color,braket,hyphenat,makeidx,xcolor,dsfont,subfigure}
\usepackage[ocgcolorlinks,colorlinks=true,linkcolor=blue,citecolor=red,linktocpage=true]{hyperref}

\newcommand{\cl}{\mathrm{c}}
\newcommand{\h}{\mathrm{h}}

\begin{document}

\title{Survival and extreme statistics of work, heat, and \\ entropy production in steady-state heat engines}

\author{Gonzalo Manzano}
\affiliation{Institute for Cross-Disciplinary Physics and Complex Systems IFISC (UIB-CSIC), Campus Universitat Illes Balears, E-07122 Palma de Mallorca, Spain.}
\affiliation{Institute for Quantum Optics and Quantum Information IQOQI, Austrian Academy of Sciences, Boltzmanngasse 3, 1090 Vienna, Austria.}

\author{\'Edgar Rold\'an} 
\affiliation{ICTP - The Abdus Salam International Centre for Theoretical Physics, Strada Costiera 11, 34151 Trieste, Italy}

\begin{abstract}
We derive universal bounds for the finite-time survival probability of the stochastic work extracted in steady-state heat engines and the stochastic heat dissipated to the environment. We also find estimates for the time-dependent thresholds that  these quantities does not surpass with a prescribed probability. At long times, the tightest thresholds are proportional to the large deviation functions of stochastic entropy production. Our results entail an extension of martingale theory for entropy production, for which we derive universal inequalities involving its maximum and minimum statistics that are valid for generic Markovian dynamics in non-equilibrium stationary states. We test our main results with numerical simulations of a stochastic photoelectric device.
\end{abstract}

\maketitle 

\section{Introduction}

The magnitude of extreme values in stochastic processes and how often these may occur is a topic of primary interest in statistical physics with multidisciplinary applications~\cite{Gumbel1958}. Extreme fluctuations, although generally associated to small probabilities, may have a critical impact in many physical systems across different scales, like seismic activity leading to earthquakes or Tsunami waves, prize fluctuations producing crashes in the stock market, or violent winds impacting the performance of wind turbine loads and the power grid~\cite{Lucarini2016}. In small systems, fluctuations play a prominent role, often  pushing systems far away from equilibrium~\cite{Jarzynski2011}. As a consequence, they are of crucial importance for the performance and robustness of microscopic motors, heat engines and refrigerators~\cite{Toyabe2011,VdB2014,rana2014single,Tasaki2016,Proesmans2016,Benenti2017,Pietzonka2018,Martinez2016,Rossnagel2016,Koski2015,Sivak2020,saha2019stochastic}, where the work extracted, the heat dissipated, and the efficiency over a finite time, become stochastic quantities that can be described within the framework of stochastic thermodynamics~\cite{Sekimoto2010,Seifert2012,peliti2021stochastic}.

Recent work~\cite{Neri2017,Chetrite2019,Manzano2019,Guillet2020} has applied the theory of Martingales~\cite{Doobs1953}~---a mathematical framework widely used in quantitative finance---  to unveil new universal statistical properties in nonequilibrium stochastic thermodynamics.   A milestone of this theory is the fact that the stochastic entropy production~\cite{Lebowitz1999,Seifert2005} $ S_\mathrm{tot}(\tau)\equiv S_\mathrm{tot}[\mathbf{X}_\tau]$ generated along single stationary trajectories $\mathbf{X}_\tau =\{X(s)\}_{s=0}^\tau$ is an exponential martingale~\cite{Neri2017,Chetrite2011}, i.e.  $\langle e^{- S_\mathrm{tot}(\tau)/k_{\rm B}} |\mathbf{X}_t \rangle = e^{-S_\mathrm{tot}(t)/k_{\rm B}}$, where $k_{\rm B}$ denotes Boltzmann's constant.
Here and further, we denote by $\langle \Omega(\tau) | \mathbf{X}_\tau\rangle$ the conditional average  of a functional $\Omega(\tau)\equiv \Omega[\mathbf{X}_\tau]$ over trajectories $\mathbf{X}_\tau$ whose history up to time $t<\tau$ is known, {  and we take, for the ease of notation, $k_B=1$} . Applying Doob's theorems~\cite{Doobs1953} for martingale processes to stochastic entropy production, universal equalities and inequalities concerning  first-passage and  minima of entropy production have been recently derived~\cite{Neri2017,Chetrite2019,Manzano2019,Singh2019,Guillet2020,Neri2019,Neri2020,Manzano2021}  and tested experimentally in nanoelectronic~\cite{Singh2019,Singh2019B,Manzano2021} and granular systems~\cite{Cheng2020}. However, it remains an open, active  area of research  to derive universal ``survival"  statistics (e.g. the probability to remain below or above a given threshold) of fluctuating physical quantities in steady-state nonequilibrium processes~\cite{Mori2020,Lacroix2020}. {  Moreover,} in the context of stochastic thermodynamics, it is relevant to study extreme and survival statistics of the work extracted by microscopic heat engines, {  of the heat dissipated into the environment, or} the peaks in the consumption of chemical fuel driving a molecular machine{,   since they can shed new light on the function and properties of such systems. In particular, some} important questions are: {  (i) are there universal bounds on the statistics of entropy production maxima and minima during a prescribed interval?} (ii) what is the survival probability for the work {  or heat} not to exceed or fall below a certain threshold value? (iii) what is the ``optimal" threshold that guarantees a prescribed value of the survival probability for the work extracted by a stationary heat engine?

\begin{figure}[t!]
 \includegraphics[width= \linewidth]{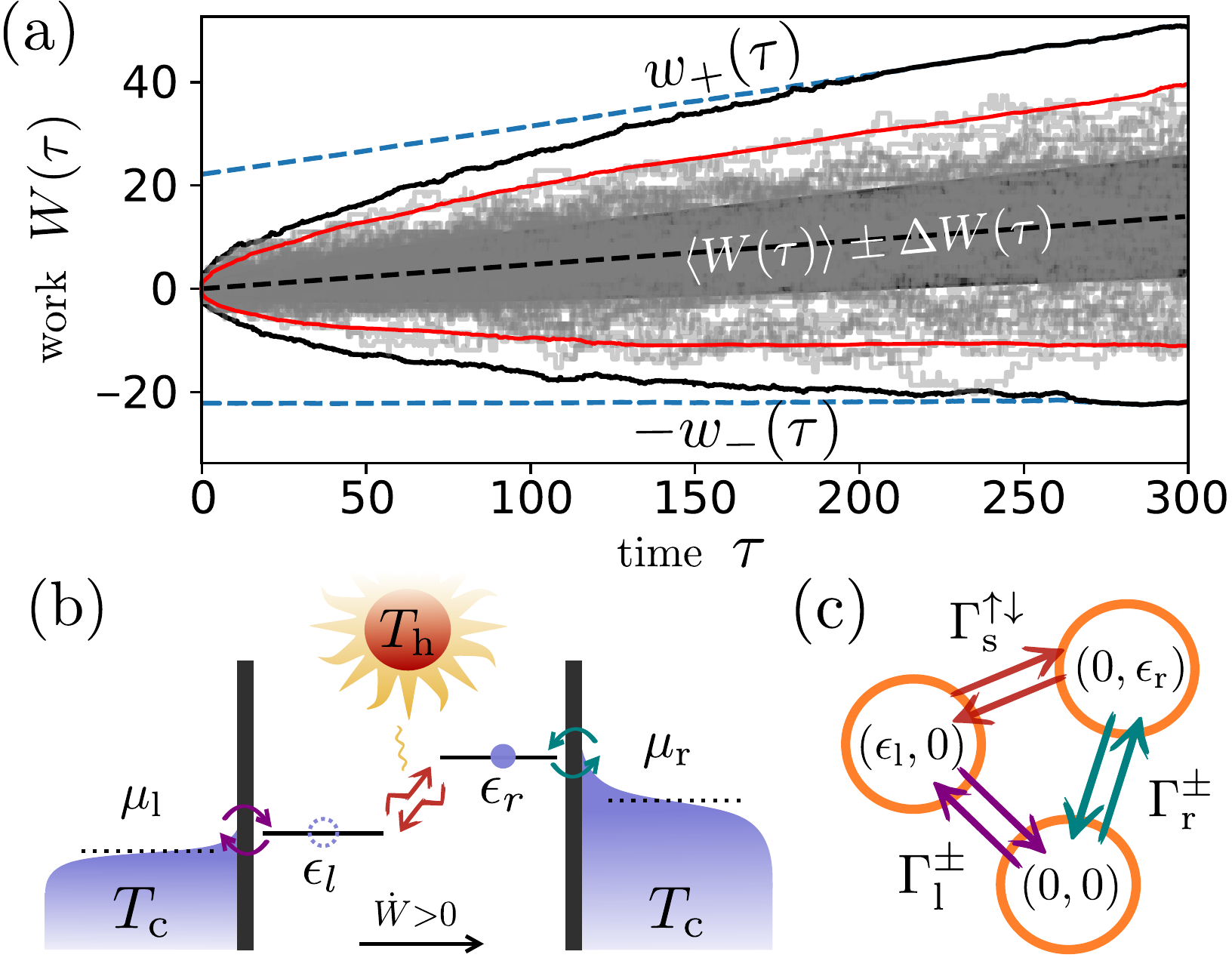}
 \caption{{\bf (a)} Illustration of work extreme fluctuations and the optimal threshold bounds $w_{\pm}(\tau)$ for its maxima (upper black and red solid lines) and minima (bottom black and red solid lines) developed by a photoelectric device during the interval $[0,\tau]$ as a function of $\tau$. Gray thin lines represent the stochastic work extracted (in $k_\mathrm{B} T$ units) for a sample of 100 trajectories. The solid lines are the optimal thresholds $w_+(\tau)$ (top) and $w_-(\tau)$ (bottom) from Eqs.~\eqref{eq:w+} and \eqref{eq:w-} respectively for a confidence value of 99\% ($\alpha = 0.01$, black lines), and   90\%  ($\alpha=0.1$, red lines). The average work output $\langle W(\tau) \rangle$ (black dashed line) and its standard deviation $\Delta W(\tau)$ (dark shadow area) are also shown for comparison. The blue dashed lines are the linear optimal asymptotic thresholds in Eq.~\eqref{eq:power} for $\alpha = 0.01$.
 {\bf{(b)}} Photoelectric device composed by two (single-level) quantum dots transporting electrons between their respective fermionic reservoirs at temperature $T_\mathrm{c}$ against a chemical potential difference ($\mu_\mathrm{r} \geq \mu_\mathrm{l}$) powered by hot photons at temperature $T_\mathrm{h}$. {\bf(c)} Energetic states of the device and relevant rates producing transitions between them (simultaneous occupancy of the two dots is avoided by Coulomb repulsion).}
 \label{fig:1}
\end{figure}

In this {  paper}, we provide insights about the above questions by first deriving universal  bounds for the cumulative distribution of the finite-time maximum and minimum of stochastic entropy production {  and their averages, substantially extending and generalizing previous results}.  We then apply these results to bound the survival statistics of the work extracted {  and the heat dissipated} by steady-state engines permanently coupled to two heat baths at cold $T_\cl$ and hot $T_\h$ temperatures. {  In particular}, we focus on the {\em survival} probability for the stochastic work extracted $W(t)$ to not surpass a threshold $w \geq 0$ anytime within the interval $t\in [0,\tau]$ (with $\tau$ sufficiently large). 
We show that the finite-time survival probability of the work can exceed  a given prescribed probability $\alpha<1$, i.e.,
\begin{equation} \label{eq:intro}
    \mathds{P}_+(w,\tau) = \mathsf{Pr}\left(\max_{0\leq t\leq \tau} W(t) < w \right) \geq 1- \alpha
\end{equation} 
whenever the threshold satisfies
\begin{equation} \label{eq:w+}
    w \geq w_+ (\tau)  = \frac{\eta T_\cl}{\eta_C - \eta} \ln\left[ \min_{p\geq1}   \frac{\langle e^{p S_\mathrm{tot}(\tau)}\rangle^{1/p}}{\alpha^{1/p}} \right].
\end{equation}
Equivalently, we show that  the right tail of the distribution of work maxima is bounded by $\mathsf{Pr}\left(\max_{t\leq \tau} W(t) > w_+(\tau) \right) \leq  \alpha$.
In other words, we find that with reliability $1 - \alpha$,  the maxima (i.e. peaks) of the  work   
are given by the right-hand side of Eq.~\eqref{eq:w+}. Analogously, we also obtain a family of lower bounds for the probability of the  minimum of the work extracted  to not fall  below a threshold $-w \leq 0$ in $t\in[0,\tau]$. Namely, we find that 
\begin{align} \label{eq:intro2}
&\mathds{P}_{-}(-w,\tau) = \mathsf{Pr}\left(\min_{t \in [0, \tau]} W(t) > -w \right) \geq 1- \alpha,
\end{align}
where the inequality is verified whenever
\begin{align} \label{eq:w-}
&w   \geq  w_-(\tau) = \frac{\eta T_\cl}{\eta_C - \eta} \ln \left[ \min_{p\geq1}  \frac{\langle e^{-p S_\mathrm{tot}(\tau)}\rangle^{1/p}}{\alpha^{1/p}} \right].
\end{align}
The \emph{optimal} thresholds $w_{\pm}(\tau)$, given by the right-hand side of  Eqs.~(\ref{eq:w+}) and~(\ref{eq:w-}), depend on the engine's efficiency $\eta$ and its distance from Carnot efficiency $\eta_C = 1- T_\cl/T_\h$, the temperature of the cold bath $T_\cl$, and the entropy-production generating function $\langle e^{\pm p S_\mathrm{tot}(\tau)}\rangle$ at the final time $\tau$. This implies that reaching Carnot efficiency $\eta \rightarrow \eta_C$ at finite dissipation, $\langle e^{\pm p S_\mathrm{tot}(\tau)}\rangle > \alpha$ leads to diverging power extreme fluctuations, as suggested by Ref.~\cite{Pietzonka2018}.  Remarkably, the $\tau-$dependent optimal thresholds depend on the desired  probability $\alpha$ of absorption.  Thus, they provide bounds for the extreme excursions of work for different lengths of the operation interval $\tau$, as illustrated  in Fig.~\ref{fig:1} for the example of a photoelectric device.

In the large operation limit, $\tau \rightarrow \infty$, we can approximate $\ln \langle e^{\pm p S_\mathrm{tot}(\tau)}\rangle^{1/p}\simeq (\tau/p) \lambda_{{S}} (\pm p)$, where $\lambda_{{S}} (k)\equiv \lim_{\tau\to\infty}(1/\tau)\ln \langle e^{k  {S}_\mathrm{tot}(\tau)}\rangle $ is the scaled cumulant generating function of entropy production \cite{Touchette2009}. Notably, in this limit, the work optimal thresholds become linear with $\tau$ and the optimization problem can be solved exactly: 
\begin{align} \label{eq:power}
w_\pm (\tau) &\simeq  \tau  \frac{\eta T_\cl}{\eta_C - \eta} \min_{p\geq 1} \left[\lambda_S(\pm p) -\frac{\ln \alpha}{\tau p}\right]\nonumber\\
&=
\frac{\eta T_\cl}{\eta_C - \eta}  \Big[\tau \lambda_S(\pm 1)  - \ln \alpha \Big],
\end{align}
where we have used the convexity of the scaled cumulant generating function and the fluctuation theorem $\lambda_S(-p)=\lambda_S(p+1)$~\cite{Lebowitz1999}.
Note that because $\langle e^{-S_{\rm tot}(\tau)}\rangle=1$ then $\lambda_S(-1) =0$ and hence the optimal lower threshold saturates to a time-independent value $w_- \simeq - (\ln \alpha) \eta T_c /(\eta_C-\eta) $. On the other hand, the optimal upper threshold reaches a linear saturating behaviour $w_+ (\tau) \simeq  \tau \lambda_S(1)  \eta T_c /(\eta_C-\eta)$ whose slope is independent on the significance level $\alpha$. In practical situations, it might be approximated using the lower bounds to the generating function of entropy production~\cite{Pietzonka2016,Gingrich2016,Polettini2016}. 

{ 
The above results can be generalized to any other integrated current proportional to the entropy production, such as the heat dissipated into the cold reservoir $Q_\mathrm{diss}(t)$ during the interval $[0, \tau]$, for which we derive bounds and thresholds analogous to Eqs.~\eqref{eq:intro}-\eqref{eq:power} limiting the fluctuations in the events of extreme heat dissipation by the engine, as we will discuss in detail later on.}

\section{Entropy production extreme fluctuations}

The theoretical results in Eq.~\eqref{eq:intro}-\eqref{eq:w-} are based on a family of universal bounds for the probability of entropy production maxima and minima which we derive and discuss in the following [Eqs.~\eqref{eq:supremum.family}-\eqref{eq:infimum.family}], together with inequalities for the averages maximum and minimum of entropy production [Eqs.~\eqref{eq:supremum.laws}].
In what follows, we consider a  system described by a discrete variable $X(t)$ which evolves in time following a Markovian, continuous-time  nonequilibrium stationary process. 
The system is assumed to be in contact with one or several heat baths at temperatures $T_k$ ($k\geq 1$) such that every transition $n \to m$ takes place at an average rate $W_k(n,m)$ due to bath $k$. Hence, the rates obey local detailed balance, i.e.  $W_k(n,m)= W_k(m,n) e^{\sigma_k(n,m)}$,  with the $\sigma_k(n,m)$ entropy change of the $k$-th bath in the $n\to m$ transition.
The stochastic entropy production associated with  a single trajectory $\mathbf{X}_\tau$ is given by:
\begin{equation} \label{eq:stot}
  S_\mathrm{tot}(\tau) = \ln \frac{P[\mathbf{X}_\tau]}{P[\tilde{\mathbf{X}}_\tau]} = \Delta S(\tau) - \sum_k \frac{Q_k(\tau)}{T_k}, 
\end{equation}
where $P[\tilde{\mathbf{X}}_\tau]$ is the probability of observing the time-reversed path $\tilde{\mathbf{X}}_\tau = \{X(\tau -t)\}_{t=0}^\tau$. Here, we have also introduced $\Delta S(t) = \ln(P_{\rm st}[X(0)]/P_{\rm st}[X(\tau)])$ the stochastic system entropy change~\cite{Seifert2005} and $Q_k(\tau) \equiv - T_k \int_0^\tau dt \sigma_k[X(t),X(t+dt)]$ the stochastic heat absorbed from the $k$-th bath along the trajectory $\mathbf{X}_\tau$. From the definition~\eqref{eq:stot}  it follows that  $e^{-S_{\rm tot}(t)}$ is a martingale. This implies that $e^{q S_\mathrm{tot}(\tau)}$  is a submartingale process  for any real $|q|\geq 1$, that is 
\begin{align} \label{eq:submartingale1}
 \langle e^{q S_\mathrm{tot}(\tau)} | \mathbf{X}_t  \rangle & \geq  e^{q S_\mathrm{tot}(t)}.
 \end{align}
 whenever  $\tau\geq t\geq 0$. Equation~\eqref{eq:submartingale1} follows from applying Jensen's inequality for conditional expectations. In particular, for the martingale $M(t)=e^{-S_{\rm tot}(t)}$ and the convex function $f(x)=x^{-q}$ ($|q|\geq 1$), we have  $\langle f(M(\tau)) | \mathbf{X}_t  \rangle \geq  M(t)$ for any  $\tau\geq t\geq 0$. Equivalently, for the choice $f(x)=-\ln x$ one gets   $\langle S_\mathrm{tot}(\tau) | \mathbf{X}_t  \rangle \geq  S_\mathrm{tot}(t)$ which generalizes the Second law $\langle S_{\rm tot}(\tau)\rangle\geq 0$.
 
 We now apply  Doob's  maximal inequality~\cite{Doobs1953,Williams} $\mathsf{Pr}\left(\mathrm{max}_{t \leq \tau} e^{q S_\mathrm{tot}(t)} \geq s \right) \leq \langle e^{q S_\mathrm{tot}(\tau)} \rangle/s$   to the positive submartingales  $e^{q S_\mathrm{tot}(\tau)}$ (recall that $|q|\geq 1$), which yields, changing variables, the following universal bounds for the cumulative distribution of entropy production maxima and minima $-$and thus for its survival probability 
\begin{subequations} \label{eqs:family}
 \begin{align} 
\label{eq:supremum.family}
 \mathsf{Pr}(S_\mathrm{max} (\tau) \geq s) &\leq  e^{-p s}\langle e^{p  S_\mathrm{tot}(\tau)} \rangle \\ \label{eq:infimum.family}
 \mathsf{Pr}(S_\mathrm{min} (\tau) \leq -  s) &\leq e^{-p s}\langle e^{-p  S_\mathrm{tot}(\tau)}\rangle, 
\end{align}
\end{subequations}
where we used the short-hand notation $S_\mathrm{max}(\tau) = \mathrm{max}_{t \leq \tau}S_\mathrm{tot} (t)$ and $S_\mathrm{min}(\tau) = \mathrm{min}_{t \leq \tau}S_\mathrm{tot} (t)$, and here  $s>0$ and $p \geq 1$. {  A detailed derivation of Eqs.~\eqref{eqs:family} is provided in appendix \ref{sec:epextrema}.} Note that, applying $p=1$ in Eq.~\eqref{eq:infimum.family}, one retrieves the  bound  $ \mathsf{Pr}(S_\mathrm{min} (\tau) \leq -  s) \leq e^{- s}$ for the entropy-production infimum statistics derived in Ref.~\cite{Neri2017}. As we will show below, the value of $p$ at which the tightest bounds are obtained is strongly dependent on $\tau$ and the statistics of $S_{\rm tot}(\tau)$.  
Moreover, Eqs.~\eqref{eq:supremum.family}-\eqref{eq:infimum.family} provide  information about the survival probability with two absorbing, symmetric barriers located at $\pm s$. The probability that $S_{\rm tot}$ escapes the interval $[-s,s]$ before time $t$ obeys $\mathsf{Pr}(\mathsf{T}_{\rm esc}\leq t)\leq \mathsf{Pr}(\mathrm{max}_{t \geq \tau}S_\mathrm{tot} (t) \geq s) + \mathsf{Pr}(\mathrm{min}_{t \geq \tau}S_\mathrm{tot} (t) \leq -s)$, which using Eqs.~\eqref{eq:supremum.family}-\eqref{eq:infimum.family} yields
\begin{equation}
    \mathsf{Pr}(\mathsf{T}_{\rm esc}\leq \tau) \leq 2 e^{-ps} \langle \cosh (pS_{\rm tot}(\tau))\rangle .\label{eq:tauesc}
\end{equation}
Equation~\eqref{eq:tauesc} reveals that the cumulative escape-time distribution from $[-s,s]$ is bounded by only the even moments of entropy production.

Another related result concerns  the average finite-time running maximum and  minimum of entropy production. To this aim, we make use of Doob's $L^p$ inequality, $\langle \mathrm{max}_{t \in [0, \tau]} |Z(t)|^p \rangle^{1/p} \leq  [p/(p-1)]~ \langle |Z(\tau)|^p \rangle^{1/p}$, where  $p>1$.
Applying this theorem to the submartingales $e^{\pm S_\mathrm{tot}(\tau)}$ (see {  appendix \ref{sec:infimumlaws}} for details) we derive the following universal bounds:
\begin{subequations} \label{eq:supremum.laws} 
\begin{align}
 \hspace{-0.2cm}\langle S_\mathrm{max} (\tau) \rangle &\leq  \ln \left[\left( \frac{p}{p-1} \right) \langle e^{p S_\mathrm{tot}(\tau)} \rangle^{1/p} \right] \\ 
 \hspace{-0.2cm}\langle S_\mathrm{min} (\tau) \rangle &\geq  -  \ln \left[\left( \frac{p}{p-1} \right)  \langle e^{-p S_\mathrm{tot}(\tau)} \rangle^{1/p} \right],
\end{align}
\end{subequations}
where here $p > 1$. Notably these inequalities generalise the infimum law $\langle S_\mathrm{min}(\tau) \rangle \geq - 1$ { (in $k_B$ units)} derived in Ref.~\cite{Neri2017}. As we will show below, the optimal bounds to the maximum and minimum of entropy production are obtained for different values of $p$ that depend crucially on $\tau$. Close to equilibrium $S_\mathrm{tot}$ becomes Gaussian and hence $\langle e^{\pm p S_\mathrm{tot}(\tau)} \rangle = 1 + (p^2 \pm p) \langle S_\mathrm{tot}(\tau) \rangle$, 
and all the bounds~(\ref{eq:supremum.laws}) only depend on the average dissipation.

\section{Steady-state heat engines} \label{sec:engines}

We now apply our generic results to the  example of Markovian steady-state heat engines that are in contact with two thermal baths.  
The baths, at inverse temperatures $\beta_h=1/T_h$ and $\beta_c=1/T_c$ and chemical potentials $\mu_h$ and $\mu_c$,   exchange energy and particles with the engine. The transition rates  are assumed to fulfill local detailed balance  $\Gamma_{i\rightarrow j}^{(h/c)} = \Gamma_{j\rightarrow i}^{(h/c)} e^{\sigma_k(i,j)}$, with   $\sigma_h(i,j) = -\beta_h [\Delta E(i,j) - \mu_h \Delta N(i,j)]$ and  $\sigma_c(i,j) = -\beta_c [\Delta E(i,j) - \mu_c \Delta N(i,j)]$ the entropy change in the hot and cold baths respectively when a transition $i\to j$ happens. Here, 
$\Delta E (i,j) = E(j) - E(i)$ is the system change in energy, and $\Delta N(i,j) $ is the change of  particle number in any of the baths.
The  work  extracted  reads $W(t) = -\int_0^\tau \sum_k \mu_k \Delta N[n(t),n(t+dt)]$. 

\subsection{Bounding extreme fluctuations of work}

Using the first law $\Delta E(\tau) + W(\tau) = Q_\h(\tau) + Q_\cl(\tau)$ inside Eq.~\eqref{eq:stot}, one finds the stochastic entropy production for this model   $S_\mathrm{tot}(\tau) = -\beta_\cl[W(t) + \Delta F_\mathrm{c}(\tau)] + (\beta_\mathrm{c} - \beta_\mathrm{h}) Q_\mathrm{h}(\tau)$, where $\Delta F_\mathrm{c}(\tau) \equiv \Delta E(\tau) - k_B T_\mathrm{c} \Delta S(\tau)$ is  a nonequilibrium free energy change. 
Note that in the stationary state,  all non-extensive quantities such as energy and entropy changes vanish on average, and all time-extensive currents become on average proportional to each other. In particular, $\langle W (\tau) \rangle= \eta \langle Q_\mathrm{h}(\tau) \rangle$ with $\eta$ the  macroscopic efficiency of the heat engine. The stochastic heat absorbed from the hot reservoir can be then approximated as $Q_\mathrm{h}(\tau) \simeq  \eta^{-1} W(\tau) + \xi(\tau)$, 
where $\xi(\tau)$ is a bounded, zero mean  $\langle \xi(\tau) \rangle = 0$ term that is non extensive in time. Under this approximation we get $S_\mathrm{tot}(\tau) = \beta_\mathrm{c} \left(\frac{\eta_C - \eta}{\eta} \right) W(\tau) + \beta_\mathrm{c}[\eta_C \xi(\tau) - \Delta F_\mathrm{c}(\tau)]$. It follows that
\begin{subequations}
\begin{align} \label{eq:wsup} 
  \max_{t \leq\tau} S_{\rm tot}(t) &\leq  \beta_\mathrm{c} \left(\frac{\eta_C - \eta}{\eta} \right) \max_{t \leq\tau} W(t) + \kappa, \\ \label{eq:winf}
\min_{t \leq\tau} S_{\rm tot}(t) &\geq  \beta_\mathrm{c} \left(\frac{\eta_C - \eta}{\eta} \right) \min_{t \leq\tau} W(t) - \kappa, 
\end{align}
\end{subequations}
where $\kappa \geq \beta_\mathrm{c} \max_{t \in [0,\tau]}[\eta_C \xi(t) - \Delta F_{\mathrm{c}}(t)] \geq 0$ is a global upper bound to a non-extensive stochastic quantity. Using inequalitites~\eqref{eq:wsup}-\eqref{eq:winf} 
in Eqs.~\eqref{eq:supremum.family}-\eqref{eq:infimum.family}, we obtain  the following bounds for the finite-time survival probability of  the  work extracted
\begin{subequations} \label{eq:survival}
\begin{align} \label{eq:survivalM}
\mathds{P}_+(w,\tau) &\geq 1 - e^{-p w \frac{\eta - \eta_C}{\eta T_\cl}} \langle e^{~p (S_\mathrm{tot}(\tau) - \kappa)} \rangle, \\ \label{eq:survivalm}
\mathds{P}_-(-w,\tau) &\geq 1 - e^{-p w \frac{\eta - \eta_C}{\eta T_\cl}} \langle e^{-p (S_\mathrm{tot}(\tau) - \kappa)} \rangle.    
\end{align}
\end{subequations}
Using Eqs.~\eqref{eq:survivalM}-\eqref{eq:survivalm} we can estimate what is the work threshold value $w_+>0$ ($-w_-<0$) that a steady-state heat engine does not surpass (fall below) in a time interval of duration $\tau$ with a prescribed probability $1 -\alpha$. They are obtained by solving $\mathds{P}_+(w_+,\tau) \geq 1 - \alpha$ and $\mathds{P}_-(-w_-,\tau) \geq 1 - \alpha$, which yield
\begin{align} \label{eq:w.th}
w_{\pm}^{(p)}(\tau) &= \left(\frac{\eta T_\cl}{\eta_C - \eta} \right) \left[\ln \frac{\langle e^{\pm p S_\mathrm{tot}(\tau)} \rangle^{1/p}}{\alpha^{1/p}}  \mp \kappa \right], \\ \label{eq:w.th-app}
&  \simeq \left(\frac{\eta T_\cl}{\eta_C - \eta} \right) \left[\ln \frac{\langle e^{\pm p S_\mathrm{tot}(\tau)} \rangle^{1/p}}{\alpha^{1/p}} \right],
\end{align}
with $p \geq 1$. Note that the threshold values in Eq.~\eqref{eq:w.th} increase monotonically with time due to the submartingale property~\eqref{eq:submartingale1}, which implies  $\langle e^{q S_\mathrm{tot}(t+dt)} \rangle  \geq  \langle e^{q S_\mathrm{tot}(t)} \rangle$, with equality when we have  $q=-1$, or when $q\neq -1$ and $S_\mathrm{tot}(t)=0$ for all $t$, that is, in equilibrium. Thus, for $\tau$  large  or $\alpha$ sufficiently small, we have $\langle e^{\pm p S_\mathrm{tot}(\tau)}\rangle \gg \alpha e^{\pm p \kappa}$ in the above equation, and the term $\kappa$ can be neglected yielding the approximate threshold given by Eq.~\eqref{eq:w.th-app}, which imply the optimal threshold values in Eqs.~\eqref{eq:w+} and~\eqref{eq:w-}~\cite{Note0}. 

\subsection{Bounding extreme heat dissipation events}

{ 
We can use an analogous approximate relation to link the heat currents from the hot and cold baths as $Q_{\h}(\tau) \simeq - Q_{\cl}(\tau)/(1-\eta) + \xi^\prime(\tau)$, where the stochastic quantity $\xi^\prime(\tau)$ is bounded and has zero mean, so that at the average level we recover $\langle Q_\h(\tau) \rangle \simeq - \langle Q_{\cl}(\tau) \rangle/(1-\eta)$. It is related to $\xi(\tau)$ in Sec.~\ref{sec:engines} as $\xi^\prime(\tau) = [\Delta E  - \eta \xi(\tau)]/(1-\eta)$. Using the above relation, the stochastic entropy production in Eq.~\eqref{eq:stot} can be conveniently rewritten as $S_\mathrm{tot}(\tau)= \Delta S(\tau) - Q_{\cl}(\tau)(\eta_C - \eta)/(1- \eta) - \beta_\h \xi^\prime(\tau)$. Then considering the maximum over the interval $[0,\tau]$ we obtain the following relations between the maximum and minimum of entropy production and the heat \emph{dissipated} into the cold reservoir, $Q^\mathrm{diss}(\tau) \equiv - Q_{\cl}(\tau)$:
\begin{subequations}
\begin{align} \label{eq:qsup}
  S_{\mathrm{max}}(\tau) &\leq \beta_\mathrm{c} \left(\frac{\eta_C - \eta}{1-\eta} \right)  Q^{\mathrm{diss}}_{\mathrm{max}}(\tau) + \kappa^\prime, \\
  \label{eq:inf}
  S_{\mathrm{min}}(\tau) &\geq \beta_\mathrm{c} \left(\frac{\eta_C - \eta}{1-\eta} \right)  Q^{\mathrm{diss}}_{\mathrm{min}}(\tau) - \kappa^\prime, 
\end{align}
\end{subequations}
with $Q^{\mathrm{diss}}_{\mathrm{max}}(\tau) = \max_{t \in [0,\tau]}[- Q_\mathrm{c}(t)]$ the maximum of the heat dissipated into the cold reservoir during the interval $[0,\tau]$, and $Q^{\mathrm{diss}}_{\mathrm{min}}(\tau) = \min_{t \in [0,\tau]}[- Q_\mathrm{c}(t)]$ the corresponding minimum (i.e. the maximum heat absorbed). Here $\kappa^\prime \geq \max_{t \in [0,\tau]}[\Delta S(t) - \beta_\mathrm{h}\xi^\prime(t)]$ bounding again the terms non-extensive in time. Following the same procedure as in Sec.~\ref{sec:engines} we obtain analogous relations for the extreme dissipation events dumping heat to the cold reservoir. These can be expressed in terms of the survival probability for the heat dissipated, namely 
\begin{subequations} \label{eq:survivalheat}
\begin{align}
 \mathds{Q}_+(q,\tau) &\equiv \mathsf{Pr}\left(\max_{0\leq t\leq \tau} Q^\mathrm{diss}(t) < q \right) \geq 1- \alpha \\
\mathds{Q}_-(-q,\tau) &\equiv \mathsf{Pr}\left(\min_{0\leq t\leq \tau} Q^\mathrm{diss}(t) > -q \right) \geq 1- \alpha
 \end{align} 
\end{subequations}
which are respectively verified when $q \geq \min_{p \geq 1} q_{\pm}^{(p)}$ for the thresholds:
\begin{subequations} \label{eq:heat.th}
\begin{align}
q_+^{(p)}(\tau) &= T_\mathrm{c} \left( \frac{1-\eta}{\eta_C - \eta} \right)\left[\log \frac{\langle e^{p S_\mathrm{tot} (\tau)} \rangle^{1/p}}{\alpha^{1/p}}  -\kappa^\prime \right] \nonumber \\ &{ \simeq T_\mathrm{c} \left( \frac{1-\eta}{\eta_C - \eta} \right)\left[\log \frac{\langle e^{p S_\mathrm{tot} (\tau)} \rangle^{1/p}}{\alpha^{1/p}} \right]}, \\
q_-^{(p)}(\tau) &= T_\mathrm{c} \left( \frac{1-\eta}{\eta_C - \eta} \right)\left[\log \frac{\langle e^{-p S_\mathrm{tot} (\tau)} \rangle^{1/p}}{\alpha^{1/p}}  +\kappa^\prime \right] \nonumber \\ &{ \simeq T_\mathrm{c} \left( \frac{1-\eta}{\eta_C - \eta} \right)\left[\log \frac{\langle e^{-p S_\mathrm{tot} (\tau)} \rangle^{1/p}}{\alpha^{1/p}}  \right]}.
\end{align}
\end{subequations}
The thresholds depend again on the failure probability $\alpha$, and guarantee a precise limit on events leading to an extreme dissipation on the cold reservoir that may lead to e.g. its overheating. We notice that $q_\pm^{(p)}(\tau)$ above diverge if $\eta \rightarrow \eta_C$ at finite power, and that the non-extensive term $\kappa^\prime$ can again be neglected when $\tau$ is large  or $\alpha$ sufficiently small, following the same arguments given above. These results are in complete analogy to the work thresholds in Eqs.~\eqref{eq:w.th} and \eqref{eq:w.th-app} and hence lead to corresponding optimal thresholds as in Eqs.~\eqref{eq:intro}-\eqref{eq:power}.
}

\section{Photoelectric device model} 
\label{sec:example}

\begin{figure*}[tbh]
 \includegraphics[width= \linewidth]{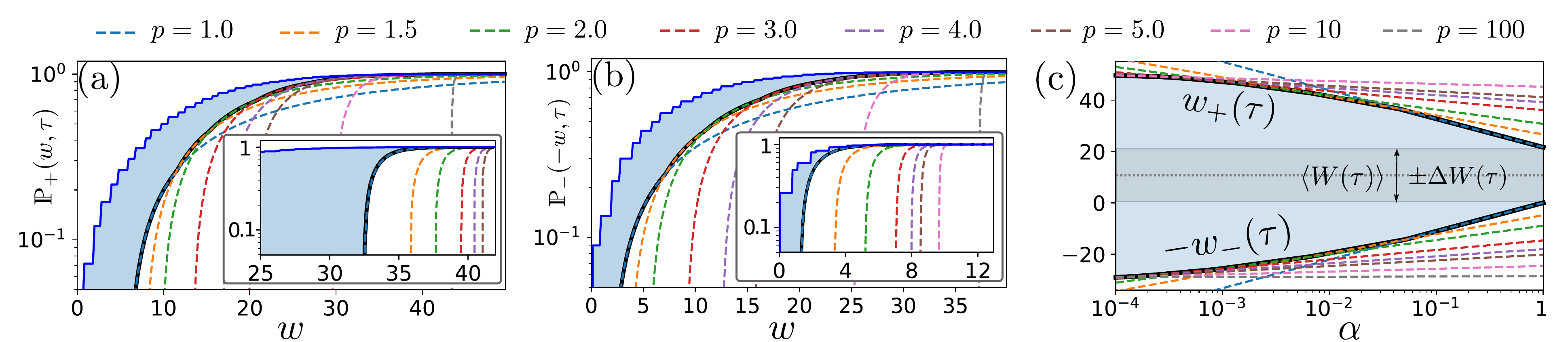}
 \caption{ {\bf{(a,b)}} Survival probability for the work extracted to not cross a given threshold [{\bf{(a)}} $w>0$, and {\bf{(b)}} $-w<0$], as a function of the threshold amplitude $w$: numerical results   (blue solid line), bounds~\eqref{eq:survivalM}-\eqref{eq:survivalm} using different values of $p$ (see top color legend), and optimal lower bounds obtained numerically (black solid line).  The main panels show results   close to the equilibrium point, $\eta= 0.95\eta_C$ ($q_\mathrm{e} V = 1.95 k_B T_\cl$) and the insets far from the equilibrium point, $\eta= 0.54 \eta_C ~(q_\mathrm{e} V = 1.1 k_B T_\cl)$,  for which  the parameter   $\kappa= 0.10$  and $\kappa= 1.12$, respectively. {\bf (c)} Value of the approximate work thresholds~\eqref{eq:w.th-app} for different $p$ (dashed lines) and optimal work thresholds given by Eqs.~\eqref{eq:w+}-\eqref{eq:w-} (black solid lines), as a function of the absorption probability $\alpha$, for $\eta= 0.83 \eta_C ~(q_\mathrm{e} V = 1.7 k_B T_\cl)$. The grey dotted line illustrates the  average output work $\langle W(\tau) \rangle$ and  the dark shaded area its standard deviation. Other simulation parameters: $T_\h=15 T_\cl, \mu_\mathrm{l}=0.8 k_B T_\cl$,  $\eta_C= 0.93$ , $\epsilon_\mathrm{l}= 1.3 k_B T_\cl$, $\epsilon_\mathrm{r}= 3.5 k_B T_\cl$, $\alpha=0.005$, $\Gamma_0 \tau = 230$, and $N\sim 10^4$ independent runs. 
 \vspace{-0.5cm}} 
 \label{fig:2}
\end{figure*}

We illustrate our results for the paradigmatic case of a stochastic photoelectric device~\cite{Cleuren2009,VdB2014,Sothmann2014}, see  Fig.~\ref{fig:1}b. Two dots  with energy levels $\epsilon_\mathrm{l}$ (left) and $\epsilon_\mathrm{r}$ (right) exchange electrons with two fermionic baths that are at the same temperature $T_\cl$, but different chemical potentials $\mu_\mathrm{l}$ and $\mu_\mathrm{r}$ respectively. A photonic reservoir temperature $T_\h$ induces transitions between the two dots, while non-radiative recombination processes are neglected~\cite{Cleuren2009,Queisser1961}.  
Due to Coulomb repulsion, the two energy levels cannot be occupied simultaneously, i.e. there are three possible configurations of the system. 
The energies of the three  states of the system are  $\epsilon_0=0$, $\epsilon_\mathrm{l}>0$, and $\epsilon_\mathrm{r}>\epsilon_\mathrm{l}$, see Fig.~\ref{fig:1}b for an illustration with the transition rates.
The tunnelling rates from the leads ($i=\mathrm{l},\mathrm{r}$) are given by $\Gamma_i^{+}=\gamma_0 f_i(\epsilon_i)$, $\Gamma_i^{-}= \gamma_0 f_i(1 -\epsilon_i)$, with $\gamma_0>0$, and $f_i(x)=1/(e^{(x-\mu_i)/T_\cl}+1)$ the corresponding Fermi distribution. For the bosonic reservoir $\Gamma_\mathrm{s}^\uparrow = \Gamma_0 \bar{n}$, $\Gamma_\mathrm{s}^\downarrow = \Gamma_0 (\bar{n} +1)$ where $\bar{n} = 1/(e^{(\epsilon_\mathrm{r}-\epsilon_\mathrm{l})/T_\h}-1)$ is the average number of photons  in the bath. For simplicity we assume in the following $\gamma_0 = \Gamma_0$. It can be straightforwardly checked that $\Gamma_i^{-} = \Gamma_i^{+} e^{-(\epsilon_i - \mu_i)/T_\mathrm{c}}$ for $i=\mathrm{l,r}$ and $\Gamma_\mathrm{s}^{\downarrow} = \Gamma_\mathrm{s}^{\uparrow} e^{-(\epsilon_\mathrm{r} - \epsilon_\mathrm{l})/T_\mathrm{h}}$. The efficiency of the device is given by the ratio between the applied voltage between leads $V = (\mu_\mathrm{r} - \mu_\mathrm{l})/q_\mathrm{e}$, with $q_\mathrm{e}$ the elementary charge, and the difference in energies of the dots, as $\eta = q_\mathrm{e} V/(\epsilon_\mathrm{r}- \epsilon_\mathrm{l})$.

We perform numerical simulations of the photoelectric device for two values of the voltage bias in the device (see insets), close and far from the equilibrium point, $V_\mathrm{\ast} = (\epsilon_\mathrm{r} - \epsilon_\mathrm{l}) \eta_C/q_\mathrm{e}$, and considered large operation times $\tau = 230~\Gamma_0^{-1}$.  
The performance of our bounds for the survival probability of work maxima $\mathds{P}_+(w,\tau)$ [Eq.~\eqref{eq:survivalM}] and minima $\mathds{P}_-(-w,\tau)$ [Eq.~\eqref{eq:survivalm}] are respectively illustrated in Fig.~\ref{fig:2}a and Fig.~\ref{fig:2}b. These bounds become tight for large $w$, corresponding to the right (left) tail of the maximum (minimum) work probability distribution. This implies that the optimal work thresholds $w_+(\tau)$ and $w_-(\tau)$ in Eqs.~\eqref{eq:intro}-\eqref{eq:w-} are tight for small values of the failure probability $\alpha$.
We also observe that close to the equilibrium point, as $w$ increases large values of $p$ provide the tightest bounds to $\mathds{P}_+(w,\tau)$ and $\mathds{P}_-(-w,\tau)$ (black solid lines). Far from equilibrium however, $p=1$ is accurate even for large $w$.

In Fig.~\ref{fig:2}c we illustrate our results for the upper and lower work thresholds in  Eq.~\eqref{eq:w.th-app} as a function of the  failure probability $\alpha$ for fixed $\tau$ and different values of $p$ (dashed lines). For this case, we empirically determine the quantity  $\kappa = 0.34$, which can be neglected in~\eqref{eq:w.th} when $\langle e^{\pm p S_\mathrm{tot}(\tau)}\rangle/ e^{\pm p \kappa} \gg \alpha$, see appendix~  \ref{app:photo}. In addition, for the case $p=1$, we need $\alpha \ll e^{-\kappa} \simeq 0.7$ to guarantee the accuracy of the lower threshold $w_{-}^{(1)}(\tau)$. 
The black solid lines show  the optimal thresholds in Eqs.~\eqref{eq:intro}-\eqref{eq:intro2} obtained by optimizing over the range of $p$ that we explored. As can be appreciated in Fig.~\ref{fig:2}c, for large values of $\alpha$, the tighter thresholds are obtained for $p=1$, while as we reduce $\alpha$ the thresholds corresponding to higher values of $p$ become the most accurate ones. As the system is pushed far away from the equilibrium point, the transition from lower to higher $p$-values becomes more abrupt and occurs at lower values of $\alpha$ (more details are given in appendix \ref{app:photo}). We also mention that, in general, high $p$ values are important to determine the optimal thresholds only for short lengths of the operation interval, while in the limit of large $\tau$ the optimal thresholds are determined by the $p=1$ case, leading respectively to a linear growth for $w_+(\tau)$ and a constant behavior in $w_{-}(\tau)$, as predicted by Eq.~\eqref{eq:power} as shown in Fig.~\ref{fig:1}.

\begin{figure}[tbh]
 \includegraphics[width= \linewidth]{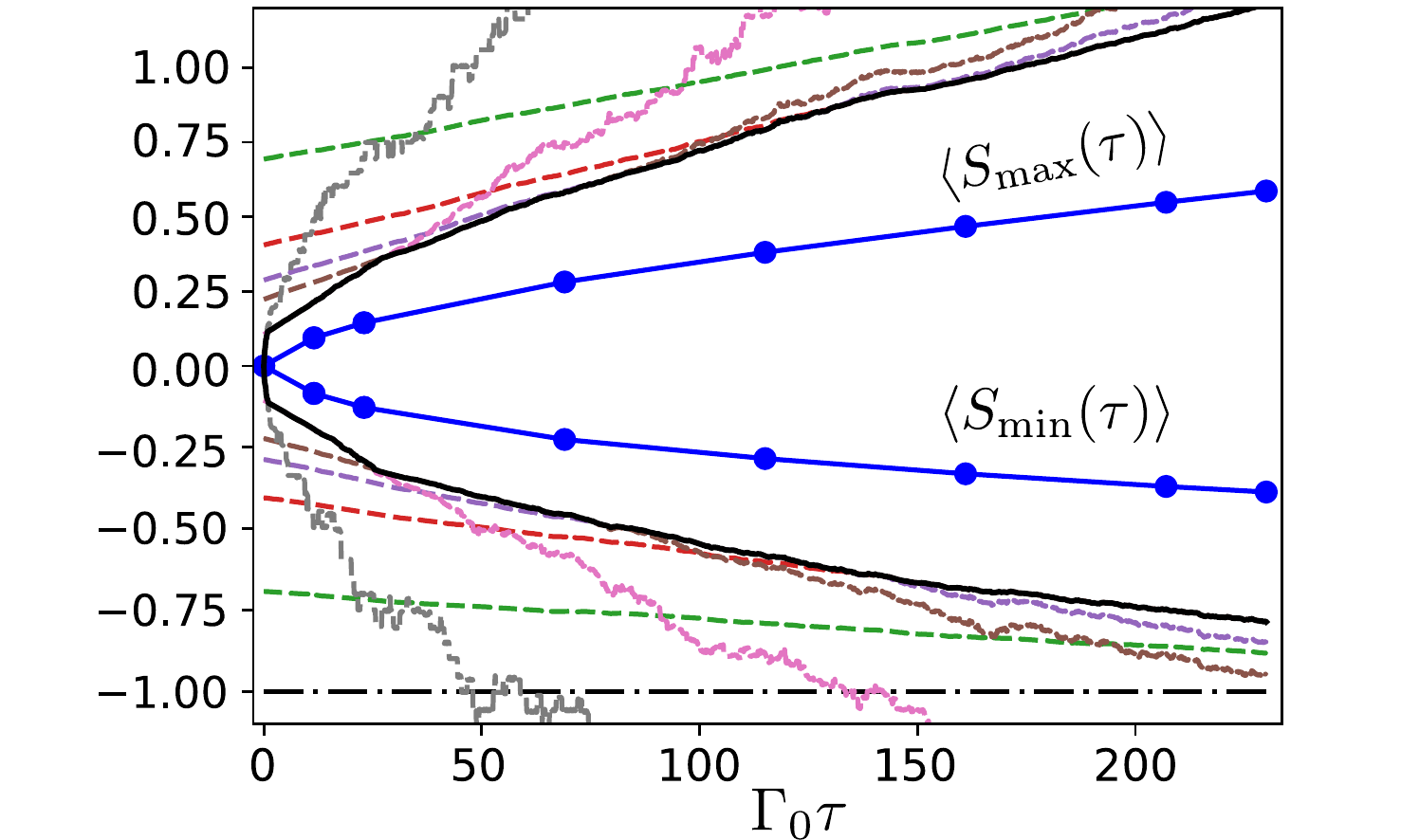}
 \caption{Average maximum and minimum of entropy production (blue circles) in $k_\mathrm{B}$ units as a function of the observation time and their respective bounds in Eq.~\eqref{eq:supremum.laws}. Optimal bounds are given by the solid black lines, while dashed lines represent different values of $p>1$, following the color code in Fig.~\ref{fig:2}. Parameters of the simulation: $T_\h=15 T_\cl, \mu_\mathrm{l}=0.8 k_B T_\cl$,$\eta= 0.95\eta_C$, $\epsilon_\mathrm{l}= 1.3 k_B T_\cl$, $\epsilon_\mathrm{r}= 3.5 k_B T$, $N\sim 10^4$ trajectories.} 
 \label{fig:3}
\end{figure}

Fig.~\ref{fig:3} illustrates the generalized supremum and infimum laws for entropy production given by  Eqs.~\eqref{eq:supremum.laws}. Large values of $p$ provide tighter upper and lower bounds for short times, where the minimum of the entropy production becomes relevant. On the other hand, small values of $p$  dominate as longer observation times  are considered,  where the maximum becomes the dominant. Moreover we observe that, for all the time interval lengths considered here, the optimal  lower bounds to $\langle S_\mathrm{min}(\tau) \rangle$ with $p \geq 2$ are tighter than the infimum law~\cite{Neri2017} $\langle S_\mathrm{min}(\tau) \rangle\geq -1$ (dashed-dotted line).

{  Finally, it is worth remarking that here we have considered, for the ease of illustration, an ideal photoelectric device without non-radiative electron transfer between the two dots. However our theory can be applied as well to more realistic situations with both radiative and non-negligible non-radiative channels, by separately identifying the jumps produced by each mechanism along the stochastic evolution.}

\section{Conclusions}


We derived several universal nonequilibrium inequalities for the entropy production extrema statistics (running maximum and minimum) that are valid in arbitrary nonequilibrium steady-states. {  This includes a family of bounds for the probability to surpass arbitrary positive and negative thresholds in a given time interval [Eqs.~\eqref{eq:supremum.family}], a bound for the distribution of scaping times [Eq.~\eqref{eq:tauesc}] and a set of inequalities for the average maximum and minimum of entropy production [Eqs.~\eqref{eq:supremum.laws}] that extend and complement the infimum law derived in Ref.~\cite{Neri2017}.

We then applied these general results to the case of steady-state heat engines that produce work while being permanently coupled to heat reservoirs at different temperatures. For this case we obtained inequalities for the survival probabilities of work [Eqs.~\eqref{eq:survival}] and heat [Eqs.~\eqref{eq:survivalheat}], together with time-dependent thresholds that allow to bound their extreme fluctuations with a given confidence level [Eqs.~\eqref{eq:intro}-\eqref{eq:power} and Eqs.~\eqref{eq:heat.th} respectively].}

Our results may be relevant in e.g. the design of suitable load systems {  that would be} connected to the engine { to profit from the work extracted, and} which need to both absorb and release the corresponding peaks of work {  while ensuring a correct operation}. Similar implications follow for the design of accurate detectors monitoring its heat exchange~\cite{Karimi2020} {  and whose operation would need to be guaranteed during a prescribed time interval without overheating.} 

{  As a final remark, we mention that while here we focused on steady-state heat engines, analogous results can be obtained for the case of power-driven and absorption refrigerators~\cite{Segal2018,Saryal2021}, as well as other thermal machines and energy-harvesting devices~\cite{Sothmann2014,Benenti2017} that operate in steady-state conditions. It would be also interesting to apply our results to the case of molecular motors powered by chemical reactions~\cite{Sivak2020,Martinez2019,Speck2021} and to explore extensions to hybrid thermal machines performing different useful thermodynamic tasks simultaneously~\cite{Manzano2020}.}
 
\begin{acknowledgments}
G. Manzano acknowledges financial support from Spanish MICINN through the program Juan de la Cierva-Incorporaci\'on  (IJC2019-039592-I) and from the European Union's Horizon 2020 research and innovation program under the Marie Sk\l{}odowska-Curie grant agreement No. 801110 and the Austrian Federal Ministry of Education, Science and Research (BMBWF).
\end{acknowledgments}

\appendix

\section{Proof of the bounds on the entropy production extrema statistics} \label{sec:epextrema}

Martingales are stochastic process $M(\tau)$ that are bounded, $|M(t)| < \infty$ for any finite time $t$, and that verify $\langle M(\tau) | \gamma_{[0,t]} \rangle = M(t)$~for $0\leq t \leq \tau$. Similarly, submartingales $Z(t)$ are bounded stochastic process that verify $\langle Z(\tau) | \gamma_{[0,t]} \rangle \geq Z(t)$~\cite{Williams}. A powerful property of non-negative martingale and submartingale processes is that they verify  Doob's submartingale inequality~\cite{Doobs1953,Williams}:
\begin{equation} \label{eq:doobs.sub}
\lambda~ \mathsf{Pr}\Big(\mathrm{max}_{t \in [0, \tau]} Z(t) \geq \lambda \Big) \leq \langle Z(\tau) \rangle, 
\end{equation}
where $\lambda > 0$ is a positive constant. The above inequality puts a constraint on the right tail of the distribution of the maximum of $Z$ along the interval $[0,\tau]$ by its mean at the final time $\tau$, and the value of the threshold $\lambda$. Another related result is Doob's $L^p$ inequality, which bounds the average of the maximum of $Z(t)$ in the interval over many trajectories $\langle \mathrm{max}_{t \in [0, \tau]} |Z(t)|^p \rangle^{1/p} \leq  q~ \langle |Z(\tau)|^p \rangle^{1/p}$, where $p^{-1} + q^{-1} = 1$ and $p>1$.

Let us consider now the finite-time maxima and minima of entropy production, $S_\mathrm{max}(\tau) \equiv \max_{t \in [0, \tau]} S_\mathrm{tot}(t)$ and $S_\mathrm{min}(\tau) \equiv \min_{t \in [0, \tau]} S_\mathrm{tot}(t)$, during the generic time interval $[0,\tau]$. We apply Doob's maximal inequality [Eq.~\eqref{eq:doobs.sub}] to $Z(t) = e^{\pm p S_\mathrm{tot}(t)}$ in Eq.~\eqref{eq:submartingale1} with $p \geq 1$ (here for the ease of notation we take $k_B = 1$). For the case $Z(t) = e^{+ p S_\mathrm{tot}(t)}$ we obtain the following family of exponential bounds for the statistics of entropy production maxima in Eq.~\eqref{eq:supremum.family}:
\begin{align} 
\label{eq:supremum.family.supp}
 \mathsf{Pr}(S_\mathrm{max} (\tau) \geq s) &\leq e^{-p s} \langle e^{p S_\mathrm{tot}(\tau)} \rangle, \nonumber
\end{align}
where $s > 0$ is a positive threshold and we made the identification $\lambda = e^{p s}$ in Eq.~\eqref{eq:doobs.sub}. Remarkably, the above equations imply that the probability of extreme (positive) entropy production events during the interval $[0, \tau]$, decreases exponentially with the threshold $s$ (and $p\geq1$), multiplied by the moment generating function of the entropy production $\langle e^{p S_\mathrm{tot}(\tau)} \rangle$ at the end of the interval. This set of bounds provide a tight envelope to the statistics of $S_\mathrm{max}(\tau)$, with higher (lower) values of $p$ providing narrow bounds for larger (smaller) values of the threshold $s$.

Analogously, the case $Z(t) = e^{- p S_\mathrm{tot}(t)}$ give us a family of lower bounds for the probability that the finite-time minimum of entropy production, $S_\mathrm{min}(\tau)$, lies above some negative threshold $-s$ (with $s > 0$). To obtain this family of bounds we notice that $S_\mathrm{min}(\tau) = - \mathrm{max}_{t \in [0, \tau]} \ln Z(t)/p = -(1/p) \ln \mathrm{max}_{t \in [0, \tau]} Z(t)$ since the logarithm is monotonically increasing. Then by using Eq.~\eqref{eq:doobs.sub} and the above chain of equalities we obtain Eq.~\eqref{eq:infimum.family}: 
\begin{equation} \label{eq:infimum.family.supp}
\mathsf{Pr}(S_\mathrm{min} \leq - s) \leq e^{-p s} \langle e^{-p S_\mathrm{tot}(\tau)} \rangle, \nonumber
\end{equation}
where again $p \geq 1$ and we identified $\lambda = e^{p s}$. This provides a family of exponential and time-dependent bounds to entropy production minima. For $p=1$ we recover from Eq.~\eqref{eq:infimum.family.supp} and the integral fluctuation theorem $\langle e^{- S_\mathrm{tot}(\tau)} \rangle = 1$, the result in Ref.~\cite{Neri2017}. The later bound is saturated in the continuous (diffusive) limit when $t \rightarrow \infty$, as reported in Refs.~\cite{Singh2019B,Cheng2020}.

\section{Proof of the generalized supremum and infimum laws for entropy production} \label{sec:infimumlaws}

The generalized maximum and infimum laws reported in Eq.~\eqref{eq:supremum.laws} can be derived by using Doob's $L^p$ inequality~\cite{Doobs1953,Williams}:
\begin{equation}
 \langle \mathrm{max}_{t \in [0, \tau]} |Z(t)|^p \rangle^{1/p} \leq  \left(\frac{p}{p-1}\right)~ \langle |Z(\tau)|^p \rangle^{1/p},
\end{equation}
where $Z(t)$ is a submartingale and $p>1$. Applying this inequality to the positive submartingales $Z(t)=e^{\pm S_\mathrm{tot}(t)}$ (again we take here $k_B = 1$) we obtain, respectively:
\begin{equation}
 \langle \mathrm{max}_{t \in [0, \tau]} e^{ \pm p S_\mathrm{tot}(t)} \rangle^{1/p} \leq  \left(\frac{p}{p-1}\right)~ \langle e^{\pm p S_\mathrm{tot}(t)} \rangle^{1/p}.
\end{equation}
We now use that since the exponential function is a monotonically increasing function of $x$ we have $\mathrm{max}_{t \in [0, \tau]} e^{ \pm p S_\mathrm{tot}(t)} = \exp(p \mathrm{max}_{t \in [0, \tau]} \pm S_\mathrm{tot}(t))$ and since $\mathrm{max}_{t \in [0, \tau]} - S_\mathrm{tot}(t) = - \mathrm{min}_{t \in [0, \tau]} S_\mathrm{tot}(t)$, we obtain the two following inequalities:
\begin{align} \label{eq:1}
 \langle  \exp[p S_\mathrm{max}(\tau)] \rangle^{1/p} &\leq  \left(\frac{p}{p-1}\right)~ \langle e^{\pm p S_\mathrm{tot}(t)} \rangle^{1/p}. \\ \label{eq:2}
\langle  \exp[- p S_\mathrm{min}(\tau)]| \rangle^{1/p} &\leq  \left(\frac{p}{p-1}\right)~ \langle e^{\pm p S_\mathrm{tot}(t)} \rangle^{1/p},
\end{align}
where we identified $S_\mathrm{max}(\tau)$ and $S_\mathrm{min}(\tau)$ respectively. Now we use in the left-hand-side of Eq.~\eqref{eq:1} that the exponential function is convex, hence fulfilling Jensen's inequality $\langle e^x \rangle \geq e^{\langle x \rangle}$, and in the left-hand-side of Eq.~\eqref{eq:2} that the function $\phi(x) = x^p$ is also convex for $x>0$ (we recall that $e^{- S_\mathrm{min}(\tau)}$ is non-negative) and hence Jensen's inequality implies $\langle |x|^p \rangle \geq |\langle x \rangle|^p$. We then obtain the following lower bounds to the right hand sides of Eqs.~\eqref{eq:1} and Eq.~\eqref{eq:2} respectively:
\begin{align} \label{eq:1p}
 \exp[ \langle S_\mathrm{max}(\tau) \rangle] &\leq  \left(\frac{p}{p-1}\right)~ \langle e^{\pm p S_\mathrm{tot}(t)} \rangle^{1/p}. \\ \label{eq:2p}
\exp[- \langle S_\mathrm{min}(\tau) \rangle] &\leq  \left(\frac{p}{p-1}\right)~ \langle e^{\pm p S_\mathrm{tot}(t)} \rangle^{1/p}.
\end{align}
Finally by taking logarithms in both sides of the equations Eq.~\eqref{eq:1p} and \eqref{eq:2p} and multiplying Eq.~\eqref{eq:2p} by $-1$ we immediately recover Eqs.~\eqref{eq:supremum.laws}.

\section{Details on the approximated work thresholds for the photoelectric device model} \label{app:photo}

In this appendix we provide additional details about the approximate time-depedent  thresholds used to bound the extreme statistics of work for the photoelectric device model employed to illustrate our main results. We first check the suitability of the approximation leading to Eq.~\eqref{eq:w.th-app} for the set of parameters used in Fig.~\ref{fig:2}. Then we provide some extra discussions about the behavior of the thresholds when shifting from close to far from equilibrium parameters.

\begin{figure*}[bth]
 \includegraphics[width= \linewidth]{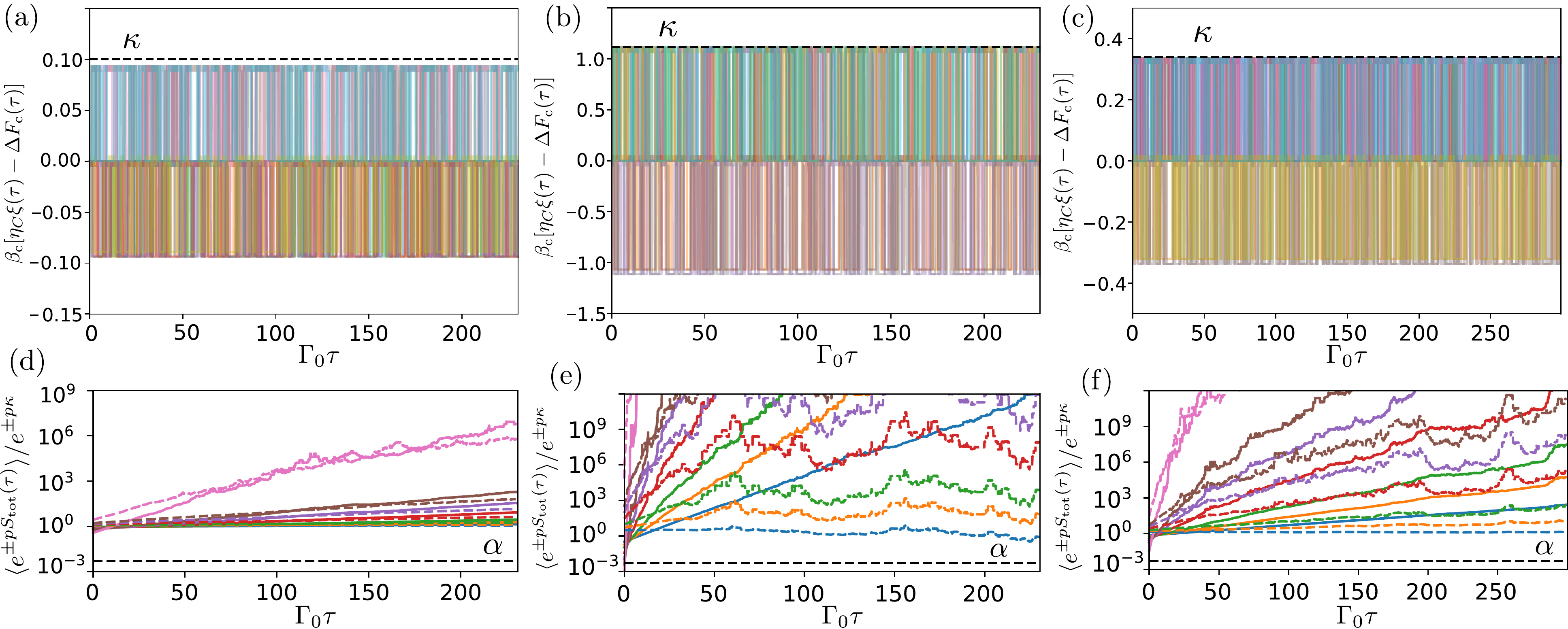}
 \caption{{\bf (a)-(c)} Sample trajectories of the non-extensive (bounded) quantity $\beta_\mathrm{c}[\eta_C \xi(t) - \Delta F_{\mathrm{c}}(t)]$ and chosen values for the parameter $\kappa$ (black dashed lines) for the three different set of parameters used: {\bf (a)} close to the equilibrium point, $\eta= 0.95\eta_C$ ($q_\mathrm{e} V = 1.95 k_B T_\cl$) leading to $\kappa = 0.10$, {\bf (b)} far from the equilibrium point, $\eta= 0.54 \eta_C ~(q_\mathrm{e} V = 1.1 k_B T_\cl)$ where $\kappa = 1.12$ and {\bf (c)} at intermediate point $\eta= 0.83 \eta_C ~(q_\mathrm{e} V = 1.7 k_B T_\cl)$ for which $\kappa= 0.34$. {\bf (d)-(f)} Testing the adequacy of the approximation leading to the final form of work thresholds in Eq.~\eqref{eq:w.th-app} for the corresponding values of $\kappa$ and parameters as in (a)-(c). We compare the ratio $\langle e^{\pm p S_\mathrm{tot}(\tau)} \rangle /e^{\pm p \kappa}$ for different values of $+ p$ (solid lines) and $-p$ (dashed lines) [see color legend in Fig.~\ref{fig:extra}] as a function of $\gamma_0 \tau$, with the chosen value of the failure probability $\alpha = 0.005$ (bottom black dashed line). Other parameters of the simulations are as in Fig.~\ref{fig:2}.} 
 \label{fig:extra2}
\end{figure*}

Let us first focus on the parameter $\kappa$ appearing in our bounds for the finite-time survival probability of the work extracted by the engine, Eqs.~\eqref{eq:survival}, used in Fig.~\ref{fig:2}a and Fig.~\ref{fig:2}b. Since $\kappa$ should upper bound the stochastic non-extensive (bounded) quantity $\beta_\mathrm{c}[\eta_C \xi(t) - \Delta F_{\mathrm{c}}(t)]$, we plot the later for a small sample of trajectories and take $\kappa$ as an upper bound to it. This is shown in Fig.~\ref{fig:extra2}(a)-(c), where the grey lines are different instances of $\beta_\mathrm{c}[\eta_C \xi(t) - \Delta F_{\mathrm{c}}(t)]$, and the chosen value of $\kappa$ is the black dashed line. As can be appreciated $\kappa$ becomes larger as we move far from the equilibrium point, see e.g. case (b), for which $\eta=0.53 \eta_C$, leading to $\kappa \simeq 1$. Nevertheless, as mentioned before one can always guarantee sufficiently large time interval $\tau$, and (or) a sufficiently small probability $\alpha$, such that $\langle e^{\pm p S_\mathrm{tot}(\tau)} \rangle \gg \alpha e^{\pm p \kappa}$ for $p \geq 1$, and hence $\kappa$ can be neglected in the threshold expressions in Eq.~\eqref{eq:w.th}, leading to the approximate thresholds \eqref{eq:w.th-app}. In Fig.~\ref{fig:extra2}(d)-(f) we compare the adequacy of such an approximation for the same three cases mentioned above. We plot the ratio $\langle e^{\pm p S_\mathrm{tot}(\tau)} \rangle /e^{\pm p \kappa}$ for $p=1,1.5,2,3,4,5$ and $10$ as compared to the values of $\alpha= 0.005$ used in Fig.~\ref{fig:2}, as a function of the time interval length $\Gamma_0 \tau$. The approximation is well justified whenever $\langle e^{\pm p S_\mathrm{tot}(\tau)} \rangle /e^{\pm p \kappa} \ll \alpha$. In the plot the solid (dashed) lines correspond to $+p$ ($-p$) cases, respectively. As can be appreciated, the ratios increase faster for increasing $p$ (specially for the cases $+p$) that would guarantee the approximation for arbitrary values of $\alpha \leq 1$, except in the case $-1$, where the ratio becomes constant in time due to the fluctuation theorem (slight variations from $1$ are due to finite statistics with $N \sim 10^4$ trajectories). In any case, a probability $\alpha \ll 1$ suffices to ensure the overall suitability of the approximation leading to Eq.~\eqref{eq:w.th-app} and the final form of the optimal thresholds in Eqs.~\eqref{eq:w+} and \eqref{eq:w-}.

\begin{figure*}[bht]
 \includegraphics[width= \linewidth]{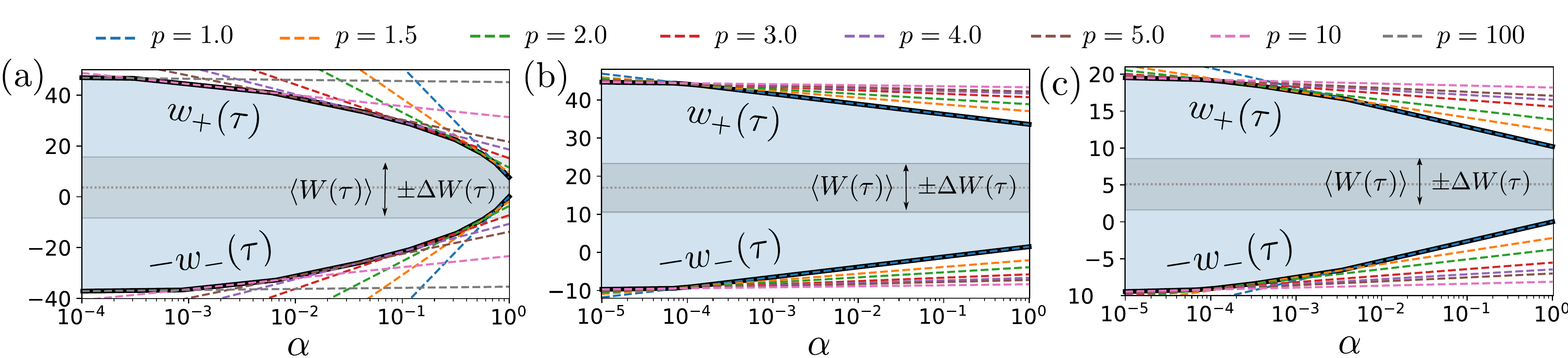}
 \caption{Work thresholds in Eq.~\eqref{eq:w.th-app} as a function of the failure probability $\alpha$ for different values of $p$ (dashed lines) according to the top color legend. The optimal thresholds in Eqs.\eqref{eq:w+} and \eqref{eq:w-} that guarantee bounded fluctuations are highlighted by the solid black lines. {\bf (a)} Close to equilibrium conditions as given by $\eta= 0.95\eta_C$ ($q_\mathrm{e} V = 1.95 k_B T_\cl$) and interval length $\Gamma_0 \tau = 230$. {\bf (b)} Far from equilibrium conditions, $\eta= 0.54 \eta_C ~(q_\mathrm{e} V = 1.1 k_B T_\cl)$ and interval length $\Gamma_0 \tau = 230$. {\bf (c)} Same far from equilibrium conditions and short time interval $\Gamma_0 \tau = 70$. In the three cases we also show the average work $\langle W(\tau) \rangle$ at the end of the interval $[0,\tau]$, together with its standard deviation $\Delta W(\tau)$. Parameters of the simulation: $T_\h=15 T_\cl, \mu_\mathrm{l}=0.8 k_B T_\cl$, $\epsilon_\mathrm{l}= 1.3 k_B T_\cl$, $\epsilon_\mathrm{r}= 3.5 k_B T$, $N\sim 10^4$ trajectories.} 
 \label{fig:extra}
\end{figure*}

As mentioned in Sec.~\ref{sec:example} a feature of the work thresholds $w_{\pm}^{(p)}$ in Eq.~\eqref{eq:w.th-app} is that in far from equilibrium situations, where the entropy production is high, there is a more abrupt transition in the optimal thresholds from the bounds given by low values of $p$ to high values. This effect can be appreciated by comparing Fig.~\ref{fig:extra}a and Fig.~\ref{fig:extra}b where we show, respectively, the family of thresholds~\eqref{eq:w.th-app} for different values of $p$ in close ($\eta = 0.95 \eta_C$) and far-from-equilibrium situations($\eta = 0.54 \eta_C$) for long intervals $\Gamma_0 \tau = 230$ as a function of the failure probability $\alpha$. As it can be appreciated there, while in the former case all  different values of $p$ play an important role for some value of $\alpha$, leading to a smooth curve, in the later situation the thresholds for $p=1$ dominate until very low values of $\alpha$ where the thresholds quickly saturate to high values, $p= 10$ (notice that in Figs.~\ref{fig:extra}b and ~\ref{fig:extra}c we do not include the more demanding case $p=100$), making intermediate values superfluous, and quickly bending the curve at $\alpha \simeq 10^{-4}$. This also implies that in far from equilibrium situations, computing the simpler case $p=1$ gives a good approximation to the optimal thresholds even for moderate and low values of the failure probability up to $\alpha\simeq 10^{-4}$. Another feature that distinguish close from far-from-equilibrium situations is that while the upper threshold $w_+(\tau)$ take similar values in both situations, substantial differences appear on the values of the lower threshold $w_{-}(\tau)$, whose values can be multiplied by a factor 4 in the far-from-equilibrium case with respect to close to equilibrium. In other words, close to equilibrium the work thresholds are more symmetric and centered around zero, indicating that high fluctuations of either work extracted or performed by the load are similar. In contrast far away from equilibrium high fluctuations for work exerted on the system get suppressed more quickly than those for work extracted. Moreover the standard deviation $\Delta W(\tau)$ of the work distribution becomes notably smaller in this case.

A similar trend as described above for the transition between work thresholds from low to high values of $p$, and on the asymmetry of upper and lower thresholds, can be also appreciated by comparing with Fig.~\ref{fig:extra}c, which corresponds to the same far-from-equilibrium conditions than Fig.~\ref{fig:extra}b ($\eta= 0.54 \eta_C$), but much smaller time-intervals, $\Gamma_0 \tau = 70$.  We also see that for shorter lengths of the time-interval the transition from low to high values of $p$ in the optimal thresholds is displaced towards higher values of $\alpha$. This implies that the simplest case $p=1$ give the optimal threshold (or a good approximation to it) only for values of $\alpha$ up to $10^{-3}$, in contrast to the $10^{-4}$ obtained for larger intervals. The later imply confidence intervals of $99.9\%$ and $99.99\%$ respectively for the work fluctuations to not surpass the thresholds.

\bibliography{refstot.bib}

\end{document}